\documentclass[12pt,a4paper]{article} 
\usepackage{amsmath}
\usepackage{amssymb} 
\usepackage{graphicx} 
\usepackage{float}
\restylefloat{table}
\begin{document}
\begin{titlepage} 
\begin{flushright} IFUP--TH/2012-14r\\ 
\end{flushright} ~
\vskip .8truecm 
\begin{center} 
\Large\bf Accessory parameters for Liouville theory on the torus 
\end{center}
\vskip 1.2truecm 
\begin{center}
{Pietro Menotti} \\ 
{\small\it Dipartimento di Fisica, Universit{\`a} di Pisa and}\\ 
{\small\it INFN, Sezione di Pisa, Largo B. Pontecorvo 3, I-56127}\\
{\small\it e-mail: menotti@df.unipi.it}\\ 
\end{center} 
\centerline{July 2012}
                
\vskip 1.2truecm
                                                              
\begin{abstract}
We give an implicit
equation for the accessory parameter on the torus which is the necessary and
sufficient condition 
to obtain the monodromy of the conformal factor.  It is shown that
the perturbative series for the accessory parameter 
in the coupling constant converges in a finite disk
and give a rigorous lower bound for the radius of convergence. We work out
explicitly the perturbative result to second order in the coupling for the
accessory parameter and to third order for the one-point function.  
Modular invariance is discussed and exploited. At the
non perturbative level it is shown that the accessory parameter is a
continuous function of the coupling in the whole physical region 
and that it is analytic except at most a finite number of
points. 
We also prove that the accessory parameter as a function of the
modulus of the torus is continuous 
and real-analytic except at most for a
zero measure set. Three soluble cases in which the solution can be expressed 
in terms of hypergeometric functions are explicitly treated.

\end{abstract}

\end{titlepage}

\eject

\section{Introduction}\label{introduction}

Liouville theory plays an important role in several fields both at the
classical and quantum level \cite{curtrightthorn,dornotto,teschner,ZZ,olesen,
jackiwpi,akerblom,gubser,nakayama}.   
Recently a renewed interest has developed due to a
conjecture \cite{AGT,gaiotto} that Liouville theory on a Riemann surface of
genus $g$ is 
related to a certain class of N = 2, 4-dimensional gauge theories and the
conjecture has been supported by extensive tests on genera 0 and 1 
\cite{AGT,drukker,alba} and proven in a class of cases \cite{hadasz1,hadasz2}. 
At the classical level the key point in
solving the theory is the determination of the accessory parameters
which 
on the sphere are
related to the semiclassical limit of the operator product expansion via the
Polyakov relation.

The determination of the accessory parameters turns out to be a highly
transcendental problem.
The mathematical literature is concentrated mainly on the limit case of
parabolic singularities i.e. punctures. On the other hand in quantum
Liouville theory, elliptic singularities which exhibit a continuum spectrum are
of most interest.

In the three point problem on the sphere the accessory parameters are
algebraically fixed by the Fuchs relations. On the other hand the four point
problem on the sphere \cite{ZZ,MV,FLNO,ferraripiatek} 
and the one point problem on the torus 
\cite{KRV,torusI,torusII,hadasz2} lead to 
differential equations with four regular singularities which are special cases
of the Heun equation. Higher number of point lead to still more complex
equations.

The Heun accessory parameter $\beta$ depends on three quantities: the coupling
$\eta$ 
whose 
physical range is $0<\eta\leq 1/2$, the modulus $\tau$ and a scale
parameter. The 
dependence of $\beta$ on the scale parameter is trivial while  $\beta$ turns
out to be 
a weight two modular form with some simple conjugation and inversion
properties. This 
allows to predict through invariance argument the value of $\beta$ in two
special cases \cite{KRV,torusI,torusII}: 1) The so called 
harmonic case i.e. the square; 2) The equianharmonic case i.e. the rhombus with
opening angle $\pi/6$. In both cases the value of $\beta$ is zero. In these
two cases the Heun equation reduces through respectively a quadratic and a
cubic transformation to an hypergeometric equation and thus the conformal
factor can be explicitly given in terms of hypergeometric functions
for any value of the source strength in the physical region
\cite{torusI,torusII}. 
Such a reduction is possible due to  a symmetry in the parameters
$e_1,e_2,e_3$ which together with infinity give the position of the
singularities in a two sheet plane which describes the torus.

In order to connect the Heun equation to more familiar cases Maier \cite{maier}
examined all rational substitution of the independent variable which transform
the Heun equation into an hypergeometric equation. He found that this
transformation occurs only for certain polynomial of degree $2, 3, 4,5, 6$ .
The harmonic case corresponds to the order two and the equianharmonic 
case to an order three case. The other transformations introduce in addition
to the 
physical source some additional ``kinematical sources'' which correspond to
spurious sources. Thus despite the interest of the transformation, as far as
the single source problem on the torus is concerned, no new physically 
interesting case is reached. 

The nature of the dependence of the accessory parameters on the moduli and
the source strengths is not completely known.  The reason is that while the
proof of the uniqueness of the solution is relatively simple
\cite{picard,poincare} 
the existence of
the solution relies on a variational method i.e. on the minimization of a
certain functional \cite{lichtenstein,troyanov}. 
In \cite{lichtenstein} this was achieved by expanding the
conformal factor in
terms of a complete set of functions while the more modern treatment of
\cite{troyanov} exploits the techniques of Sobolev spaces proving first the
existence of a weak solution and then the existence of the solution.
The outcome is that it is very difficult at the end to follow the nature of
the dependence of the solution on the coupling and the moduli.   

An exception is the case of parabolic singularities (punctures) 
were general
properties of fuchsian mappings can be applied. Using such a technique
Keen, Rauch and Vaquez \cite{KRV} found
that the accessory parameter for the torus with one parabolic singularity
(puncture) is a real-analytic functions of the modulus; 
in addition in \cite{KRV} some numerical
investigation of the accessory parameter was performed. Zograf and Takhtajan
\cite{ZT} treated the case of parabolic singularities on
a Riemann surface of genus $0$. 
The result of \cite{KRV,ZT} is 
that the accessory parameters are real-analytic functions of the moduli. 
Kra \cite{kra} gave an extension of such a
result to the case of a collection of parabolic and 
a special class of elliptic singularities i.e. finite order elliptic
singularities where the strength of the source can assume only the values 
$\eta =1/2(1-1/n),~~n\in Z_+$.  This is a discrete set which
accumulates to the parabolic point. On the other hand in quantum
Liouville theory, elliptic singularities which exhibit a continuum spectrum are
of most interest.

In the general case of elliptic singularities and parabolic singularities it
was proved in \cite{CMS1,CMS2} that the accessory 
parameters are real-analytic in the couplings and in the moduli in an
everywhere dense open set: given a value of the coupling, if the accessory
parameter is not analytic at that point there is an open set as near as we
like to the given point, on which the accessory parameter is analytic.

Here we shall prove a much stronger result i.e. that the accessory parameter 
for
the torus is an analytic function of the coupling in the whole physical region
except at most a finite number of points and it is a real-analytic function of
the modulus in the whole fundamental region except a zero measure set.
 
In proving such results we shall rely on some properties of the 
solution which are extracted using potential-theory techniques which were used
in the solution of the uniformization problem, combined with some results on
analytic varieties \cite{whitney}. The first
is the existence and uniqueness property of the solution, a result which goes
back to Picard himself. The second is the boundedness property of the solution
$\phi$ of the Liouville equation and its first and second derivatives with
respect to the argument, in any region 
which excludes finite disks around the singularities and which was proved in
\cite{CMS2}.

The accessory parameter $\beta$ obeys an implicit equation. From such implicit
equation a power series expansion for $\beta$ in the coupling
$\eta$ can be extracted and we prove 
such expansion to be rigorously convergent in a finite disk. We also
compute a rigorous lower bound on the convergence radius. We compute also
explicitly the expansion of $\beta$ in $\eta$ up to second order in terms of
integrals of elliptic and related functions.

For general couplings
i.e. couplings not necessarily small, 
exploiting the uniqueness theorem and some results on
complex analytic varieties \cite{whitney} we are able to prove that the
$\beta$ which solves the monodromy problem is analytic in the whole physical
range of the coupling except at most for  a finite number of points.

The nature of the dependence of the accessory parameters on the
moduli of a punctured Riemann surface is important in several respects; e.g.
the $C^1$  nature of such a dependence 
is an essential input in proving Polyakov relation on the sphere
\cite{CMS1,CMS2,ZT,TZ}.
Here we prove that both for elliptic and parabolic singularity the
dependence of $\beta$ on the modulus is real-analytic except for a zero
measure set thus extending the results of \cite{CMS1,CMS2}. 
The technique developed here can be applied to the four or
higher point functions on the sphere and also to higher genus surfaces.

Within the AGT \cite{AGT,gaiotto}
correspondence Ferrari and Piatek \cite{ferraripiatek}
exploited the relation between 
the semiclassical limit of quantum Liouville theory and the
Nekrasov-Shatashvili limit of the $N=2$, $U(2)$ super Yang-Mills theory to give
an expression of the accessory parameter for the 4-point function on the
sphere in terms of a contour integral containing the ratio of the column
length of critical Young diagrams.

It should be possible to extend such technology to the case of the torus. On
the other hand once this is accomplished, a direct comparison with the result
obtained here will not be 
straightforward as they are based on different expansions. In the present
paper, the accessory parameter has been considered as a function of the source
strength and an expansion in the source strength given.  Instead in the
approach of \cite{ferraripiatek} an expansion of the accessory parameter in
the position
$x$ of the fourth singularity w.r.t. the position of the first $z=0$
singularity  
appears. A similar
approach which computes the accessory parameter expanding in $x$ is found in
\cite{ZZ}. As for the torus the modulus is related to the positions of the
singularities in the $u$-plane, it appears that the such expansion should
correspond to a perturbation around the degenerate case in which two
singularities coincide i.e. the infinite strip which we treat in section
\ref{solublecases}. In \cite{torusI} a general perturbation technique under
the 
variation of the moduli has been developed, and one could apply it to the
present situation allowing a direct comparison.

The paper is organized as follows: In section \ref{differentialequation} 
we write down the differential
equation in the cut-$u$ plane and derive the monodromicity condition.  The
fulfillment of a single complex implicit equation is necessary and 
sufficient to assure the monodromic behavior at all 
singularities. In section \ref{monodromies} we give the explicit expression of
the monodromy 
matrices. In section \ref{modularinvariance} 
we discuss modular invariance and the consequent determination
of the accessory parameter in two soluble cases. Section \ref{actions} 
is devoted to the
the relation of the action in different coordinate systems. Section
\ref{solublecases} gives
the exact expression of the conformal factor and of the one point function for
three soluble cases, one of which 
is the limit case of the infinite strip.  
In section
\ref{secondperturbation}  we develop
perturbation theory in the coupling constant up to the second order and give
lower bounds for its convergence radius. In section \ref{phiperturbation} 
we give a different
approach to perturbation theory by working directly with the conformal
factor. Here we are able to go easily to third order even if the control of
the convergence property of the series relies on the results of section 
\ref{secondperturbation}.  In
section \ref{analyticproperties} we derive the general analytic
properties of the accessory parameter 
both in the coupling and in the modulus. In section \ref{conclusions}
we give some concluding remark and point to some open problems.

\bigskip

\section{The differential equation and the monodromy
conditions}\label{differentialequation}

The equation
\begin{equation}\label{liouvilleeq}
-\partial_z\partial_{\bar z}\phi+e^{\phi}= 2\pi \eta\delta^2(z-z_t)
\end{equation}
does not contain information about the torus. They have to be put in through
periodic boundary conditions. 

To have a faithful representation of the torus we have to use the
two-sheet representation of the torus in the variable $u=\wp(z)$.
For simplicity and without loosing generality due to translational invariance
we shall set in this section $z_t=0$.

We recall \cite{torusI,torusII} that the problem of finding a solution to
eq.(\ref{liouvilleeq}) can be reduced to finding the value of the accessory 
parameter
$\beta$ and and of a real parameter $\kappa$, such that the expression
\begin{equation}
e^{-\varphi/2} =\frac{1}{\sqrt{2}|w_{12}|}
\big[\kappa^{-2}\overline{y_1(u)} y_1(u)-
\kappa^2 \overline{y_2(u)} y_2(u)\big]
\end{equation}
is monodromic. $y_1,y_2$ are two solutions of an ordinary differential equation
which contains the parameter $\beta$ and $w_{12}=y_1 y_2'-y_1' y_2$ 
is the constant Wronskian.

Such equation in $u$ is given by \cite{torusI}
\begin{equation}\label{udiffequation}
y''+Q y=y''+(Q_0+q) y=0
\end{equation}
where
\begin{eqnarray}
Q_0(u)&=&\frac{3}{16}\left(\frac{1}{(u-e_1)^2}+ \frac{1}{(u-e_2)^2}
+\frac{1}{(u-e_3)^2}+\frac{2 e_1}{(e_1-e_2)(e_3-e_1)(u-e_1)}\right.\nonumber\\
&+& \left.\frac{2 e_2}{(e_2-e_3)(e_1-e_2)(u-e_2)} 
+\frac{2 e_3}{(e_3-e_1)(e_2-e_3)(u-e_3)}\right)\nonumber
\end{eqnarray}
and
\begin{equation}
q(u) = \frac{1-\lambda^2}{16}\frac{u+\beta}{(u-e_1)(u-e_2)(u-e_3)}=
\frac{1-\lambda^2}{4}\frac{u+\beta}{[\wp'(z)]^2}\equiv
\epsilon\frac{u+\beta}{[\wp'(z)]^2}~.
\end{equation}
where $e_k=\wp(\omega_k)$.

We recall that $\lambda = 1-2\eta$ and that $\eta$ has to satisfy the Picard
condition $0<\eta\leq \frac{1}{2}$, the lower limit being due to the negative
nature of the curvature in the bulk and the upper to the local finiteness of
the area. 
The upper limit correspond to a puncture, or parabolic
singularity. The range of $\epsilon$ is $0<\epsilon\leq\frac{1}{4}$.

We  know two independent solution to eq. 
\begin{equation}
y''+Q_0 y=0~.
\end{equation}
They are
\begin{equation}
y^{(0)}_1 = [4(u-e_1)(u-e_1)(u-e_1)]^{1/4}
= [\wp'(z)]^{1/2}\equiv \Pi(u),~~
\end{equation}
\begin{equation}
y^{(0)}_2 = (z-\omega_3) [4(u-e_1)(u-e_1)(u-e_1)]^{1/4}\equiv
Z~[\wp'(z)]^{1/2}= Z~\Pi(u)~.
\end{equation}
Defining
\begin{equation}
K(u,u')=y_1^{(0)}(u)\theta(u,u')\frac{q(u')}{w_{12}}y_2^{(0)}(u')-
y_2^{(0)}(u)\theta(u,u')\frac{q(u')}{w_{12}}y_1^{(0)}(u')
\end{equation}
we can solve the full equation (\ref{udiffequation}) by the convergent
expansion 
\begin{equation}
y_j = (1+ K+KK+KKK+\dots)y^{(0)}_j~.
\end{equation}
It will be useful to write
\begin{equation}
f_j(z)=\frac{y_j(u)}{\Pi(u)} 
\end{equation}
and now the $f_j(z)$ are given by
\begin{equation}
f_1(z)=1+\epsilon \int^z_{\omega_3} 
(z'-z)(\beta+\wp(z'))dz'+
\end{equation}
\begin{equation}
+\epsilon^2
\int^z_{\omega_3} (z'-z)(\beta+\wp(z'))dz'\int^{z'}_{\omega_3} 
(z''-z')(\beta+\wp(z''))dz''+ \dots\nonumber
\end{equation}
and
\begin{equation}
f_2(z)=Z+\epsilon \int^z_{\omega_3} 
(z'-z)(\beta+\wp(z'))Z'dz'+
\end{equation}
\begin{equation}
+\epsilon^2
\int^z_{\omega_3} (z'-z)(\beta+\wp(z'))dz'\int^{z'}_{\omega_3} 
(z''-z')(\beta+\wp(z''))Z''dz''+ \dots\nonumber
\end{equation}
with $\omega_3=\omega_1+\omega_2$.
We shall be interested in the values of $f_j$ and their
derivatives $g_j\equiv f_j'$ at the
points $\omega_1$, $\omega_2$ and $\omega_2-\omega_1$, $\omega_1-\omega_2$. 
Due to the triangular structure of the multiple integrals and the 
boundedness of $\wp(z)$ along the integration paths $\omega_3-r\omega_1$ 
and $\omega_3-r\omega_2$  we see that such $f_j$ and $g_j$ are analytic
function of $\epsilon$ and $\beta$ as the series converges absolutely for any
given $\epsilon$ and $\beta$. This is a well known fact.

We have
\begin{equation}
g_1(\omega_k) = -\epsilon
\int_{\omega_3}^{\omega_k}(\beta+\wp(z')) f_1(z') dz'
\end{equation}
and
\begin{equation}
g_2(\omega_k) = 1-\epsilon
\int_{\omega_3}^{\omega_k}(\beta+\wp(z')) f_2(z') dz'
\end{equation}
and similarly for the other values of the argument. The $f_j(\omega_k)$ and
the $g_j(\omega_k)$ are holomorphic functions of $\epsilon$ and $\beta$. 

There are two ways to impose monodromy. The first \cite{torusI}
is to exploit the symmetry for $z_t=0$ of the equation (\ref{liouvilleeq}) 
and of its solution under the
inversion $z\rightarrow -z$. As proven in \cite{torusI} it is the sufficient
to impose monodromy around the three kinematical singularities $e_j=
\wp(\omega_j)$ under a full turn in $u$ which corresponds to half turn in $z$.
A more general method is to impose monodromy under two independent cycles of
the torus \cite{torusII}. While the first method requires the solution of the
differential equation on the tracts 
$(\omega_3,\omega_1)$, $(\omega_3, \omega_2)$ the second method requires 
the solution on the longer tracts $(\omega_3, \omega_3 - 2 \omega_1)$, 
$(\omega_3, \omega_3 - 2\omega_2)$, i.e. along the full cycles. 

In order to derive non perturbative results we shall need some results from the
potential theory approach to the Liouville equation. The input we shall
use is Picard result \cite{picard}
about the existence and uniqueness of the solution 
of the uniformization problem of which eq.(\ref{liouvilleeq}) is a particular 
case.
Picard was concerned only with elliptic singularities i.e. $\eta<\frac{1}{2}$.
Later the treatment was extended to elliptic and\slash or parabolic
singularities in \cite{poincare,lichtenstein}. 
For a more modern treatment using Sobolev spaces see \cite{troyanov}. 
 
In \cite{torusI} after choosing the canonical pair of solutions $y_l$
it was proven in the first approach that the monodromies at $e_1,e_2$ have the
form
\begin{equation}\label{Ml}
M(\omega_j)=
\begin{pmatrix}
i(a_jd_j+b_jc_j)&-2ia_jb_j\\
2ic_jd_j&- i(a_jd_j+b_jc_j)
\end{pmatrix}
\end{equation}
with $a_j,b_j,c_j,d_j$ elements of a $SL(2,C)$ matrix.
We have still at our disposal a scale transformation on the canonical
solutions under which the matrices go over to
\begin{equation}
\begin{pmatrix}
i(a_jd_j+b_jc_j)&-2ia_jb_j\kappa^{-2}\\
2ic_jd_j\kappa^2&- i(a_jd_j+b_jc_j)
\end{pmatrix}.
\end{equation}
Picard existence theorem tells us that for each $j$, $a_jb_j$ and $c_jd_j$ are
either both zero or both different from zero.
From this remark it follows that necessary and sufficient condition for the 
existence of a $\kappa$ which renders 
\begin{equation}\label{offdiagonal}
a_j b_j \kappa^{-2}=\bar c_j \bar d_j \bar\kappa^2
\end{equation} 
is
that in (\ref{Ml})
\begin{equation}\label{mastersphere}
M_{12}(\omega_1)\overline{M_{21}(\omega_2)}=
M_{12}(\omega_2)\overline{M_{21}(\omega_1)}~.
\end{equation}
It is not difficult to prove \cite{torusI} that once relation
(\ref{offdiagonal}) 
is satisfied it follows that all the monodromies become $SU(1,1)$ i.e. the
conformal factor $\varphi$ becomes single valued and regular. Thus $\beta$ has
to be chosen as to satisfy (\ref{mastersphere}).

Similar considerations hold in the cycle approach \cite{torusII} 
where the necessary and sufficient condition for the realization of the
monodromic solution takes the form 
\begin{equation}\label{mastercycle}
M_{12}(C_1)\overline{M_{21}(C_2)}= 
M_{12}(C_2)\overline{M_{21}(C_1)}~.
\end{equation}

\bigskip

\section{Computation of the monodromies}\label{monodromies}

Given the complex
\begin{equation}
Y(u)=
\begin{pmatrix}
y_1(u) \\
y_2(u) 
\end{pmatrix}
\end{equation}
the monodromy matrices are defined by
\begin{equation}
\tilde Y(u) = M Y(u)
\end{equation}
where $\tilde Y(u)$ denotes the complex after a complete turn in $u$ at $e_1$
or 
$e_2$, in the first approach, or after a cycle, in the second approach.
We have also
\begin{equation}
\tilde Y'(u) = M Y'(u)
\end{equation}
and thus
\begin{equation}
\begin{pmatrix}
\tilde y_1 &\tilde y'_1\\
\tilde y_2 &\tilde y'_2
\end{pmatrix}= M
\begin{pmatrix}
y_1&y'_1\\
y_2& y'_2
\end{pmatrix}
\end{equation}
from which
\begin{equation}\label{Mexplicit}
M = 
\begin{pmatrix}
\tilde y_1 &\tilde y'_1\\
\tilde y_2 &\tilde y'_2
\end{pmatrix}
\begin{pmatrix}
y'_2&-y'_1\\
-y_2& y_1
\end{pmatrix}
\end{equation}
due to $w_{12}=y_1 y'_2 - y'_1 y_2=1$. Application of eq.(\ref{Mexplicit})
to eq.(\ref{offdiagonal}) gives
\begin{equation}
M_{12}(\omega_k)= -\tilde y_1y'_1+ \tilde y'_1y_1= 
- 2 e^{i\frac{\pi}{2}}f_1(\omega_k)g_1(\omega_k)
\end{equation}
and
\begin{equation}
M_{21}(\omega_k)= \tilde y_2y'_2-\tilde y'_2y_2= 
2 e^{i\frac{\pi}{2}} f_2(\omega_k)g_2(\omega_k).
\nonumber
\end{equation}
Thus equation (\ref{mastersphere}) becomes in the first approach 
\begin{equation}\label{explicitsphere}
f_1(\omega_1) g_1(\omega_1)
\overline{f_2(\omega_2)}\overline{g_2(\omega_2)}=
f_1(\omega_2) g_1(\omega_2)
\overline{f_2(\omega_1)}\overline{g_2(\omega_1)}.
\end{equation}
Using the cycle method the monodromy equation (\ref{mastercycle}) becomes
\begin{equation}\label{explicitcycle}
g_1(\omega_3-2 \omega_1)\overline{f_2(\omega_3-2\omega_2)}=
g_1(\omega_3-2\omega_2) \overline{f_2(\omega_3-2 \omega_1)}.
\end{equation}
Due to the uniqueness of the Picard solution the two equations are equivalent.
We shall come back to this property in section \ref{secondperturbation}.

\bigskip

\section{Modular invariance}\label{modularinvariance}

We recall some simple properties of the accessory parameter $\beta$ which are
derived from the differential equation and the uniqueness theorem.

The equation 
\begin{equation}
f''(z)+\epsilon (\wp(z)+\beta)f(z)=0
\end{equation}
has invariance properties related to the transformation properties of $\wp$
under dilatations, conjugation and modular transformations \cite{KRV}.
From 
\begin{equation}
\wp(\mu z|\mu\omega_1,\mu\omega_2)=\frac{1}{\mu^2}\wp(
z|\omega_1,\omega_2)
\end{equation} 
with $\mu\in C$ we have $\mu^2\beta(\mu\omega_1,\mu\omega_2)=
\beta(\omega_1,\omega_2)$. From $\wp(z|\bar\omega_1,\bar\omega_2)=
\overline{\wp(\bar z|\omega_1,\omega_2)}$ one obtains 
$\beta(\bar\omega_1,\bar\omega_2)= 
\overline{\beta(\omega_1,\omega_2)}$. Moreover as the lattice is left
invariant under $\omega_1\rightarrow -\omega_1$ and $\omega_2\rightarrow
-\omega_2$ and under $\omega_1\leftrightarrow \omega_2$, $\wp(z)$ is unchanged
and also $\beta$ is unchanged. Similarly $\beta$ is unchanged under
$\omega_1\rightarrow \omega_1,~ \omega_2\rightarrow \omega_2+\omega_1$. The
two transformations $\omega_1\rightarrow \omega_1,\omega_2\rightarrow 
\omega_2+\omega_1$ and $\omega_1\rightarrow \omega_2,
\omega_2\rightarrow-\omega_1$ are, apart for a dilatation, the generators $T$
and $S$ of the
modular group \cite{serre}. Thus defining $\beta[\omega_2] =
\beta(1,\omega_2)$ we can synthesize the 
transformation properties of $\beta$ as
\begin{equation}
\beta\bigg[\frac{a\tau+b}{c\tau+d}\bigg] =(c\tau+d)^2\beta[\tau]
\end{equation} 
telling us that $\beta$ is a modular form of weight $2$ and
\begin{equation}
\overline{\beta[\tau]} =\beta[-\bar\tau].
\end{equation}
From such transformation properties a few simple facts follow
\cite{KRV,torusI,torusII}. 

1) For ${\rm Re}~\tau =0$, $\beta$ is real; this describes the rectangle.

2) For ${\rm Re}~\tau = \pm\frac{1}{2}$ we have $\beta={\rm real}$.

From the fact that the stabilizer in the fundamental region is $\pm I$ except
$S$ for $\tau=i$, $ST$ for $\tau = e^{2\pi i /3}$ and $TS$ for $\tau = e^{\pi
i /3}$ \cite{serre} we have 

3) For $\tau=i$ we have $\beta=0$; this describes the square.

4) For $\tau=e^{2\pi i/3}$ we also have $\beta=0$; this describes the so
called equianharmonic case where the fundamental region is a rhombus with
opening angle $2\pi/6$. 

5) For $\tau=e^{\pi i/3}$ we also have $\beta=0$; this again describes the
equianharmonic case where the rhombus with opening angle $2\pi/6$ has a
different orientation so it does not differ from case 4.

In \cite{torusI,torusII} the explicit form of the conformal factor in terms of
hypergeometric functions was given for the cases 3 and 4,5.

From the viewpoint of the differential equation in $u$, modular transformations
boil down to a simple permutation of the $e_k$ and a scale
transformation. Thus in studying the monodromies in the $u$ cut-plane with the
first method we have a simple interchange of $e_k$ in the basic equations. If
instead we exploit the cycle approach, modular invariance is due to the group
composition properties for the transfer matrices, when we add to a given cycle
one or more cycles. This will be relevant in discussing the modular invariance
of the perturbation theory results.

\bigskip

\section{The action in different coordinates}\label{actions}

It is well known \cite{ZZ,takhtajan} that also the classical action has to be
regularized due to the logarithmic divergences which arises from the kinetic
term at the singularities. 

In this section we shall write the relation among the two regularized
action $S_z$ and $S_u$ related to the $z$- and $u$-representation of
the torus.

In the $z$-representation the action is given by
\begin{equation}
\frac{S_z}{2\pi}=\frac{1}{2\pi}\int_{D_\varepsilon}(\frac{1}{2}d\phi\wedge \bar
d\phi +e^\phi dz\wedge d\bar z)\frac{i}{2}-
\frac{\eta}{4\pi i}\oint_{\partial D_{\varepsilon}}\phi(\frac{dz}{z-z_t}-
\frac{d\bar z}{\bar z-\bar z_t})-\eta^2\log\varepsilon^2
\end{equation}
where $D_\varepsilon$ is exterior of a circle of radius $\varepsilon$ around
the source at $z_t$.
Writing
\begin{equation}
\phi = -2\eta\log|z-z_t|^2+X+ o(z-z_t)
\end{equation}
we have also
\begin{equation}\label{Sz}
\frac{S_z}{2\pi}=\frac{1}{2\pi}\int_{D_\varepsilon}(\frac{1}{2}d\phi\wedge \bar
d\phi +e^\phi dz\wedge d\bar z)\frac{i}{2}-\eta X+\eta^2 \log\varepsilon^2
\end{equation}
and the important relation
\begin{equation}\label{Xrelation}
\frac{1}{2\pi}\frac{\partial S_z}{\partial \eta} = - X.
\end{equation}
In order to compute the action explicitly in the soluble cases it is however
better to put the source at the origin $z_t=0$, which gives rise to a 
singularity
in $u$ at infinity.
The transition to the $u$-representation is given by $u=\wp(z)$
\begin{equation}\label{phitovarphi}
e^\phi dz\wedge d\bar z = e^\varphi du\wedge d\bar u
\end{equation}
i.e.
\begin{equation}
\phi = \varphi -\log J\bar J
\end{equation}
with
\begin{equation}
J = \frac{d z}{d u}= \frac{1}{\wp(z)'}~.
\end{equation}
Thus
\begin{equation}\label{phivarphi}
\phi = \varphi + \log |4(u-e_1)(u-e_2)(u-e_3)|
\end{equation}
We have
\begin{equation}
\phi = -2\eta\log|z|^2+ X + o(z)
\end{equation}
and taking into account that $u=\wp(z) =
\frac{1}{z^2}+o(z)$ we have at infinity in $u$
\begin{equation} 
\varphi = (\eta-\frac{3}{2})\log u\bar u +X^u_\infty
\end{equation} 
with
\begin{equation}\label{XuX0}
X^u_\infty = X -\log 4~.
\end{equation}
The regularized $S_u$ action now takes the form
\begin{eqnarray}\label{Su}
\frac{1}{2\pi}S_u &=& \frac{1}{2\pi}\int_{D} (\frac{1}{2}
d\varphi\wedge\bar d \varphi+e^\varphi du\wedge d\bar u)\frac{i}{2}\nonumber\\
&-&\frac{1}{8\pi i}(\eta-\frac{3}{2})
\oint_{R_u}\varphi(\frac{du}{u}-\frac{d\bar u}{\bar u})
-\frac{1}{16\pi i}\oint_{\varepsilon_k}\varphi
(\frac{du}{u-e_k}-\frac{d\bar u}{\bar u-\bar e_k})
\nonumber\\
&+&\frac{1}{2}(\eta-\frac{3}{2})^2
\log R^2_u-\frac{1}{8}\log\varepsilon_k^2=
\nonumber\\
&=&\frac{1}{2\pi }\int_D(\frac{1}{2}d\varphi\wedge\bar d\varphi+
e^\varphi du\wedge d\bar u)\frac{i}{2}
\nonumber\\
&-&\frac{1}{2}(\eta-\frac{3}{2})^2\log R^2_u-(\eta-\frac{3}{2})
X^u_\infty
+\frac{1}{8}\log\varepsilon_k^2-\frac{1}{2}X_k^u
\end{eqnarray}
where the integration in $u$ is extended to the two sheets which describe
the torus. $D$ excludes disks of radius $\varepsilon_k$ around $e_k$ and on
both sheets the exterior of a circle of radius $R_u$.

We also have
\begin{equation}
\frac{1}{2\pi}\frac{\partial S_u}{\partial \eta} = - X^u_\infty.
\end{equation}

Using eqs.(\ref{Sz},\ref{phitovarphi},\ref{Su}) we find for the relation 
between the two actions
\begin{equation}
S_z = S_u -\frac{1}{4}\log|(e_1-e_2)(e_2-e_3)(e_3-e_1)|^2-2\eta \log 2~.
\end{equation}
consistent with eq.(\ref{XuX0}).
\bigskip

\section{Soluble cases}\label{solublecases}

In this section we shall give the explicit value of the action, i.e. of the
semiclassical $1$-point function, for three
soluble cases i.e. the square, the equianharmonic case i.e. a rhombus with
opening angle $2\pi/6$ and the limit case of the infinite strip for any
coupling in the physical region.

\bigskip

1. The harmonic case: the square 

\bigskip

In \cite{torusI,torusII} the Liouville field for the harmonic case i.e. the
square was computed in
terms of hypergeometric functions. The result was with $\lambda =1-2\eta$
\begin{eqnarray}\label{mphihalfsquare}
&-&\frac{\phi(z)}{2} = - \log(\sqrt{2}|\kappa|^2)\\
&+&\log\bigg[\bigg| {_2F_1}(\frac{1-\lambda}{8},\frac{1+\lambda}{8};
\frac{3}{4};u^2(z))\bigg|^2-
|\kappa|^4 |u(z)| 
\bigg| {_2F_1}(\frac{3-\lambda}{8},\frac{3+\lambda}{8};
\frac{5}{4};u^2(z))\bigg|^2
\bigg]\nonumber
\end{eqnarray}
with
\begin{equation}
|\kappa|^4 =\frac{8}{\gamma^2(\frac{1}{4})\gamma(\frac{1-\lambda}{4})
\gamma(\frac{1+\lambda}{4})}
\end{equation}
where as usual
\begin{equation}
\gamma(x) = \frac{\Gamma(x)}{\Gamma(1-x)}~.
\end{equation}
From (\ref{mphihalfsquare}) 
we obtain
\begin{equation}
X = (3+4\eta)\log 2 -
2\log\gamma(\frac{1-2\eta}{4})+\log\gamma(1-\eta) 
\end{equation}
\bigskip
and using (\ref{Xrelation}) we have with
\begin{equation}\label{F(x)}
F(x) = \int_{\frac{1}{2}}^x \log\gamma(x')~dx'
\end{equation}

\begin{equation}\label{squareaction}
S_z({\rm square}) = -(3\eta+2\eta^2)\log 2 - 4 F(\frac{1-2\eta}{4})+F(1-\eta)+
4 F(\frac{1}{4})-F(1)~. 
\end{equation}

\bigskip

2. The equianharmonic case 

We have \cite{torusII} \footnote{We correct for a factor 2 in the argument of
the first logarithm}
\begin{eqnarray}
-\frac{\phi(z)}{2}&=&-\log(2\sqrt{2}|\kappa|^2)\\
&+&\log\left[
\left|_2F_1(\frac{1-\lambda}{12},\frac{1+\lambda}{12};\frac{2}{3};u^3(z))
\right|^2- |\kappa|^4 |u(z)^2|
\left|_2F_1(\frac{5-\lambda}{12},\frac{5+\lambda}{12};
\frac{4}{3};u^3(z))\right|^{2}
\right]\nonumber
\end{eqnarray}
with
\begin{equation}\label{kappa4rhombus}
|\kappa|^4 =9~\frac{\pi \Gamma(\frac{5-\lambda}{6})\Gamma(\frac{5+\lambda}{6})}
{\Gamma^2(\frac{1}{6})\Gamma(\frac{1-\lambda}{6})\Gamma(\frac{1+\lambda}{6})}
\end{equation}
which gives
\begin{equation}
X = 2 \log 3+\frac{7}{3}\log 2+\frac{4}{3}\eta\log 2-
\log\gamma(\frac{2}{3}+\frac{\eta}{3}) -2\log\gamma(\frac{1-2\eta}{6}) +
\log\gamma(1-\frac{\eta}{3}) 
\end{equation}
Integrating
\begin{eqnarray}\label{equirhombusaction}
S_z({\rm equianharmonic}) = -&\bigg[&
2\eta \log 3+\frac{7\eta}{3}\log 2+\frac{2\eta^2}{3}\log
2-3F(\frac{2+\eta}{3})+ 
6F(\frac{1-2\eta}{6})\nonumber\\
&-&3F(1-\frac{\eta}{3})+3F(\frac{2}{3})-6F(\frac{1}{6})+3F(1)\bigg]~.
\end{eqnarray}

\bigskip

3. The infinite strip

\bigskip

We discuss here a limit case of the torus topology which is soluble i.e. the
infinite strip. The vertical infinite strip is reached with the parameters
$e_1=2a$, $e_2=e_3 = -a$. For $a=1$ we have
\begin{equation}\label{stripQ}
Q(u) = \frac{1-\lambda^2}{16}\frac{u+\beta}{(u+1)^2(u-2)}+
\frac{3}{16}\frac{u^2-2u+9}{(u-2)^2(u+1)^2}~.
\end{equation}
The accessory parameter $\beta$ has to be fixed to $1$ otherwise the pole of
order $2$ at $u=-1$ would not have the correct kinematical value $1/4$ as it
is required for the limit of an infinite rectangle. It is of interest that the
value $1$ is already given by first order perturbation theory \cite{torusI}
\begin{equation}
\beta =\frac{\zeta(\omega_2)\bar \omega_1-
\zeta(\omega_1)\bar \omega_2}{\omega_2\bar\omega_1-\omega_1\bar\omega_2}~.
\end{equation}
In fact for the case at hand i.e. $e_1=2a$,  $e_2=e_3 = -a$ the Weierstrass
$\wp$ and 
$\zeta$ functions become \cite{batemanII}
\begin{equation}
u=\wp(z)=-a+\frac{3a}{\sin^2(\sqrt{3a}~z)},~~~~~~~~
\zeta(z) = a z +\sqrt{3a}~\frac{\cos(\sqrt{3a}~z)}{\sin(\sqrt{3a}~z)}
\end{equation}
with
\begin{equation}
\omega_1 =\frac{\pi}{2\sqrt{3a}}
\end{equation}
and thus
\begin{equation}\label{betastrip}
\lim_{\omega_2\rightarrow i \infty}~\beta = a=1~.
\end{equation}
Two independent solutions of the 
differential equation (\ref{udiffequation}) 
with $Q$ given
by  (\ref{stripQ}), canonical at $u=2$ are 
\begin{equation}
y_1 =(u-2)^{1/4}(u+1)^{1/2} ~{_2F_1}(\frac{1-\lambda}{4},\frac{1+\lambda}{4},
\frac{1}{2};\frac{2-u}{3})
\end{equation}
\begin{equation}
y_2 =(u-2)^{3/4}(u+1)^{1/2} ~{_2F_1}(\frac{3-\lambda}{4},\frac{3+\lambda}{4},
\frac{3}{2};\frac{2-u}{3})
\end{equation}
giving for the $\phi$
\begin{eqnarray}
&-&\frac{\phi}{2}= - \log(3\sqrt 2) - \log \kappa^2 \\
&+&\log\bigg[\bigg|{_2F_1}(\frac{1-\lambda}{4},\frac{1+\lambda}{4},
\frac{1}{2};\frac{2-u}{3})\bigg|^2
-\kappa^4~
|u-2|\bigg|{_2F_1}(\frac{3-\lambda}{4},\frac{3+\lambda}{4},
\frac{3}{2};\frac{2-u}{3})\bigg|^2\bigg]\nonumber
\end{eqnarray}
with
\begin{equation}
\kappa^4=\frac{1}{3}\bigg[2~\gamma(\frac{3-\lambda}{4})
\gamma(\frac{3+\lambda}{4})\bigg]^2~.
\end{equation}
We find
\begin{equation}
X = 2 \eta \log\frac{4}{3} + \log 6
+2\log \gamma(1-\eta)-2\log \gamma(\frac{1}2-\eta)
\end{equation}
from which
\begin{equation}
S_z({\rm strip}) = -\bigg[\eta^2 \log\frac{4}{3}+\eta \log 6 - 2 F(1-\eta) + 
2 F(\frac{1}{2}-\eta) + 2 F(1) - 2 F(\frac{1}{2})\bigg]~.
\end{equation}

\bigskip

\section{Second order perturbation theory
and convergence radius}\label{secondperturbation}

In this section we shall develop the perturbation theory around
$\epsilon=0$. We shall show that the perturbative series is convergent in a
neighborhood of $\epsilon=0$ and give a rigorous lower bound on the
convergence radius. We shall also give the explicit expression of the first
and second order term for the accessory parameter $\beta$. We stress
that the treatment of this section requires neither Picard's
existence and uniqueness theorem nor other results from the potential theory
approach to the problem.

The perturbative series for $\beta$ is obtained by solving the implicit
equation (\ref{explicitsphere}) or (\ref{explicitcycle}). We shall use
(\ref{explicitcycle})
\begin{equation}\label{explicitcycle2}
\epsilon F(\beta,\bar\beta,\epsilon) =
g_1(\omega_3-2 \omega_1)\overline{f_2(\omega_3-2\omega_2)}-
g_1(\omega_3-2\omega_2) \overline{f_2(\omega_3-2
\omega_1)}=0~.
\end{equation}
We notice that  
\begin{equation}\label{g1eps}
g_1(\omega_3-2 \omega_k) = 2\epsilon(\omega_k\beta-\zeta(\omega_k))   
+O(\epsilon^2)
\end{equation}
being $\zeta(z)$ the Weierstrass zeta-function, while
\begin{equation}\label{f2eps0}
f_2(\omega_3-2 \omega_k) = -2\omega_k+O(\epsilon)~.
\end{equation}
Thus after dividing (\ref{explicitcycle}) by $\epsilon$ we have for
the Jacobian,  at $\epsilon=0$ 
\begin{equation}
J=\frac{\partial (F,\bar F)}{\partial(\beta,\bar\beta)}
\bigg\vert_{\epsilon=0}= 
16|\bar\omega_2\omega_1- \omega_2\bar\omega_1|^2\neq 0~.
\end{equation}
Due to the analyticity of $F$ in $\beta,\bar\beta$ and $\epsilon$ we have that
$\beta$ will be an analytic function of $\epsilon$ in an open neighborhood of
$\epsilon=0$ \cite{gunningrossi}. 
We can  therefore develop a perturbative series around
$\epsilon=0$. 
From (\ref{g1eps},\ref{f2eps0}) we have \cite{torusI}
\begin{equation}\label{betafirst}
\beta_1 = \frac{\bar\omega_2\zeta(\omega_1)-\bar\omega_1\zeta(\omega_2)}
{\bar\omega_2\omega_1-\bar\omega_1\omega_2}~.
\end{equation}
On equation (\ref{betafirst}) we can check already the following 
properties: 1. For
$\omega_2= i\omega_1$ i.e. the square $\beta_1=0$. 2. For $\omega_2 = e^{2\pi
i/3}\omega_1$ i.e. the 
equianharmonic case $\beta_1=0$. For general $\omega_1,\omega_2$ we verify
modular invariance, i.e. invariance under the two generating transformations
\begin{eqnarray}
\omega_1 \rightarrow \omega_1,~~~~\omega_2 \rightarrow \omega_2+\omega_1
~~~~(T)
\nonumber\\
\omega_1 \rightarrow -\omega_2,~~~~\omega_2 \rightarrow \omega_1~~~~(S)~.
\end{eqnarray}
All these properties are proven easily using
\begin{equation}
\zeta(\omega_1+\omega_2) = \zeta(\omega_1)+\zeta(\omega_2). 
\end{equation}
Moreover in the limit of the infinite strip we have $\beta\rightarrow 1$ (see
eq.(\ref{betastrip})).

We come now to the second order. Developing eq.(\ref{explicitcycle2}) we
obtain  
for the accessory parameter $\beta$ to the second order
\begin{eqnarray}\label{betasecond}
\beta&=&\beta_1
-\frac{\epsilon}{\omega_1\bar\omega_2-\bar\omega_1\omega_2}
\bigg[
\bar\omega_2^2
\overline{\zeta(\omega_1)}\zeta(\omega_1)
-\zeta^2(\omega_1)
(\bar\omega_1\omega_2+\omega_1\bar\omega_2)
\nonumber\\
&-&2\bar\omega_1\zeta(\omega_1)
\zeta_2(\omega_1,\omega_3)-
2\overline{\zeta_3(\omega_1,\omega_3)}\zeta(\omega_1)
+\bar\omega_1 I(\omega_1,\omega_3)+
\nonumber\\
&+&\beta_1\bigg(-\bar\omega_2^2\omega_1\overline{\zeta(\omega_1)}
+2\omega_1\overline{\zeta_3(\omega_1,\omega_3)}+2\bar\omega_1
\zeta_3(\omega_1,\omega_3)
+\bar\omega_2\omega_1^2(2\zeta(\omega_1)+\zeta(\omega_2))\bigg)
\nonumber\\
&+&\frac{2}{3}\bar\beta_1 \bar\omega_2^3 \zeta(\omega_1)
+\frac{2}{3} \beta_1^2
\bar\omega_1\omega^3_2+\frac{2}{3}\beta_1\bar\beta_1\bar\omega^3_1\omega_2
-\{\omega_1\leftrightarrow \omega_2\} \bigg]
\end{eqnarray}
where in the above expression  $\beta_1$ is the first order result
(\ref{betafirst})
and
\begin{eqnarray}
\zeta_2(z,\omega_3)&=&\int_{\omega_3}^z \zeta(z') dz'=
\log\frac{\sigma(z)}{\sigma(\omega_3)},~~~~
\zeta_3(z,\omega_3)=\int_{\omega_3}^z \zeta_2(z',\omega_3) dz'\nonumber\\
I(z,\omega_3)&=&
\int_{\omega_3}^z \zeta^2(z') dz'~.
\end{eqnarray}
Some comments are in order about such a result. $\zeta(z)$ is an entire
function and $I(z,\omega_3)$ is a single-valued function of $z$ 
in the fundamental parallelogram due to the
absence of the constant term in the expansion of $\zeta(z)$
\begin{equation}
\zeta(z) = \frac{1}{z}-\frac{g_2z^3}{2^2\cdot 3\cdot 5}+\cdots
\end{equation}
On the other hand $\zeta_2(z,\omega_3)$ and $\zeta_3(z,\omega_3)$ are not
single-valued functions; nonetheless the combinations
\begin{eqnarray}
\bar z
\zeta_2(z,\omega_3)+
\overline{\zeta_3(z,\omega_3)}\nonumber\\
z\overline{\zeta_3(z,\omega_3)}+\bar z
\zeta_3(z,\omega_3)
\end{eqnarray}
are single valued in the fundamental parallelogram. 
In fact we have
\begin{equation}
\zeta_2(z,\omega_3) =\log z+ {\rm regular~terms}
\end{equation}
and
\begin{equation}
\zeta_3(z,\omega_3) =z\log z + {\rm regular~terms}
\end{equation}
from which it follows that the terms $\bar\omega_1\zeta_2(\omega_1,\omega_3)+
\overline{\zeta_3(\omega_1,\omega_3)}$ and
$\omega_1\overline{\zeta_3(\omega_1,\omega_3)}+\bar\omega_1 
\zeta_3(\omega_1,\omega_3)$ in (\ref{betasecond}) are well defined. 

It is very important that the paths
chosen in evaluating $\zeta_2(\omega_1,\omega_3)$ and 
$\zeta_3(\omega_1,\omega_3)$ are the same even if there is no preferred path.

One can easily verify that, as
expected, for the square and the equianharmonic case such second order
contribution 
vanishes. On the other hand is very cumbersome to verify directly the modular
invariance of eq.(\ref{betasecond}). Modular invariance of 
eq.(\ref{betasecond}) is
assured at the exact and also perturbative level by the group composition
properties of the monodromies over cycles.     
Obviously the same result (\ref{betasecond}) is
obtained using the first approach to the monodromy problem i.e. starting from
eq.(\ref{explicitsphere}). Iterating the
process in eq.(\ref{explicitcycle2}) one can go to higher orders.

We come now to the convergence radius of the perturbative series in
$\epsilon$. A rigorous lower bound on the convergence radius can be obtained
applying Rouch\'e theorem \cite{titchmarsh}.

Equation (\ref{explicitcycle2})
can be rewritten as
\begin{eqnarray}
\beta - \beta_1 +\epsilon G(\beta,\bar\beta,\epsilon) =0
\end{eqnarray}
with $\beta_1$ given by (\ref{betafirst}).
It will be useful to exploit the polarization technique \cite{dangelo} i.e. 
to introduce in addition to $\beta$ an other independent complex variable
$\beta_c$ and consider the system
\begin{eqnarray}
\beta - \beta_1 +\epsilon G(\beta,\beta_c,\epsilon) =0\label{pert}\\
\beta_c - \bar\beta_1 +\epsilon \bar G(\beta_c,\beta,\epsilon) =0\label{cpert}
\end{eqnarray}
where $\bar G$ is the analytic function obtained by conjugating in the power
expansion the coefficients of $G$. Obviously if $\beta,\beta_c$ for real
$\epsilon$ is a solution of the above system, also $\bar\beta_c,\bar\beta$ 
is a solution. If, always for real $\epsilon$, the solution is unique then we
have $\beta_c=\bar\beta$ and such solution is the solution of the monodromy
problem.

Given a positive
constant $B$ we can always find a $\delta$ such that  for $|\epsilon|<\delta$ 
\begin{equation}\label{Bbound}
B>|\epsilon \bar G(\beta_c,\beta,\epsilon)|
\end{equation}
for all $\beta,\beta_c$ with $|\beta-\beta_1|\leq B$, 
$|\beta_c-\bar\beta_1|\leq B$.
Then due to the analyticity of $G$ we can apply Rouch\'e theorem to conclude
that (\ref{cpert}) for $|\epsilon|<\delta$ has one and only one solution
$\beta_c=\beta_c(\beta,\epsilon)$, with $|\beta_c-\bar\beta_1|\leq B$. 
Moreover such
$\beta_c$ will be an analytic function of $\beta$ and $\epsilon$. We
substitute now such $\beta_c(\beta,\epsilon)$ into (\ref{pert}) where we can
again apply Rouch\'e theorem and thus find a unique solution
$\beta=\beta(\epsilon)$. For real $\epsilon$, $\beta(\epsilon),~ 
\beta_c=\bar\beta(\epsilon)$ is the unique solution of the system and being 
self
conjugate it is the solution of the monodromy problem and $\delta$ will be a
rigorous lower bound for the convergence of the perturbative
expansion. Obviously if we want to optimize the outcome, we have to choose $B$
as to render $\delta$ as large as possible. 

As we shall see in the following
choosing a too large $B$ makes the bounds on $G$ increase faster than $B$ and
thus $\delta$ has to decrease to satisfy (\ref{Bbound}). On the other hand it
is obvious that $B$ small requires $\delta$ small.

As already mentioned the first approach of eq.(\ref{explicitsphere}) 
requires the
integration along a shorter path in the $z$-plane. As the simple bounds 
on $f_j$ and $g_j$ we
shall give below behave exponentially in the length of the integration path,
the first approach, even if eq.(\ref{explicitsphere}) is more complicated than
eq.(\ref{explicitcycle}), is more apt to give a larger lower bound on
the convergence radius.

To compute such lower bound on the convergence radius we use the
following simple rigorous inequalities
\begin{equation}\label{expboundfirst} 
|f_1(\omega_1)-1|\leq \cosh(\sqrt{|\epsilon|(|\beta|+m_1)}~|\omega_2|)-1
\end{equation} 
\begin{equation} 
|f_2(\omega_1)+\omega_2|
\leq \frac{\sinh(\sqrt{|\epsilon|(|\beta|+m_1)}~|\omega_2|)}
{\sqrt{|\epsilon|(|\beta|+m_1)}}-|\omega_2|
\end{equation} 
\begin{eqnarray} 
& &|g_1(\omega_1)-\epsilon(\beta\omega_2-\zeta(\omega_2))|\\
&\leq& \epsilon ~\sqrt{|\epsilon|(|\beta|+m_1)}| ~\omega_2|~
\big[\sinh(\sqrt{|\epsilon|(|\beta|+m_1)}~|\omega_2|)-
\sqrt{|\epsilon|(|\beta|+m_1)}~|\omega_2|\big]\nonumber 
\end{eqnarray} 
\begin{equation}\label{expboundlast}  
|g_2(\omega_1)-1|\leq \cosh(\sqrt{|\epsilon|(|\beta|+m_1)}~|\omega_2|)-1
\end{equation} 
where $m_1$ is the maximum of the modulus of $\wp(z)$ along the segment
$[\omega_3,\omega_1]$, and similar inequalities for the functions $f_k$
and $g_k$ with argument $\omega_2$. More elaborate inequalities can
provide a larger lower bound for the convergence radius. We report
in Table 1 the lower bounds on the convergence radius for a few values of the
modulus 
$\tau$ obtained with the above described method. As expected the square (for
which we know that $\beta$ is zero) gives the largest lower bound. Due to the
exponential behavior of the inequalities (\ref{expboundfirst}-
\ref{expboundlast}) the bound shrinks 
to zero in the highly asymmetric configurations. 
The method applies for any $\tau$.

\begin{table}[h]
\begin{center}
    \begin{tabular}{ | l | l | l | l | l | p{1.2cm} |}
    \hline
    $\tau$ & $i$ & $2i$ & $3i$ & $4i$ & $5i$\\ \hline
    $\epsilon_c$ & $0.1202$ & 0.05244 & 0.02581& 0.01512& 0.00988\\ \hline
    $B$ & $2.1667$ & 3.80021  & 5.64317 & 7.435  & 9.19403\\ \hline
     \end{tabular}
\end{center}
\caption{$\epsilon_c$ is the rigorous lower bound on the convergence
radius. The 
physical region for the coupling is $0<\epsilon\leq 1/4$.
$B$ is the parameter appearing in eq.(\ref{Bbound})}
\end{table}

\section{$\phi$-perturbation theory}\label{phiperturbation}

In this section we shall develop perturbation theory directly from the
Liouville equation. The analytic nature of the perturbative expansion has to be
borrowed form the rigorous treatment of the previous section; on the other
hand one can easily obtain in this way the value of the action to third order
in $\eta$.

The Green function on the torus of half-periods $\omega_1={\rm real}$,
$\omega_2$,
$\tau = \omega_2/\omega_1$ is given by \footnote{
Here we use for $\theta_1$ the convention of \cite{DLMF} not the one of
\cite{batemanII}} 
\begin{equation}\label{greenfunction}
G(z) = \frac{1}{4\pi}\log
\bigg\vert\frac{\theta_1(\frac{\pi z}{2\omega_1}|\tau)}{\eta_D}\bigg\vert^2+
\frac{i(z-\bar z)^2}{16\omega_1(\omega_2-\bar\omega_2)}
\end{equation}
satisfying
\begin{equation}
\Delta G(z) = \delta^2(z) -\frac{1}{a}
\end{equation}
with $a = 4\omega_1^2 \tau_I$. The arbitrary additive constant in $G$ has been
chosen in (\ref{greenfunction}) as to have 
\begin{equation}\label{intG=0}
\int G(z) dz\wedge d\bar z=0~.
\end{equation}

Then we expand
$\phi=\phi_0+\phi_1+\phi_2+\dots$ to have
\begin{equation}\label{expphi0}
e^{\phi_0} = \frac{2\pi\eta}{a}
\end{equation}
\begin{equation}
-\frac{1}{4}\Delta\phi_1(z) = 2\pi\eta \big(\delta^2(z)-\frac{1}{a}\big)
\end{equation}
from which
\begin{equation}\label{phi1}
\phi_1(z) = -8\pi\eta G(z)~.
\end{equation}
This is true if the constant term in $G$ is chosen as in (\ref{greenfunction},
\ref{intG=0}). Next we have
\begin{equation}
-\frac{1}{4}\Delta\phi_2(z) = \frac{16\pi^2\eta^2}{a}G(z)
\end{equation}
and then
\begin{equation}
\phi_2(z) = -\frac{64\pi^2\eta^2}{a}\int G(z-z')G(z') dz'\wedge d\bar z' 
\frac{i}{2}~.
\end{equation}
Using (\ref{expphi0},\ref{phi1}) and $\theta_1'(0) = 2 \eta^3_D$, 
being $\eta_D$
the Dedekind modular function, we obtain for the $X$ of section \ref{actions}
\begin{equation}\label{X01}
X = \log\frac{2\pi\eta}{4\omega_1^2\tau_I}-
4\eta\log\bigg|\frac{\pi\eta_D^2}{\omega_1}\bigg|
\end{equation}
from which
\begin{equation}\label{action2}
\frac{1}{2\pi}S_z = \eta - \eta\log\frac{2\pi\eta}{4\omega_1^2\tau_I}+
2\eta^2\log\bigg|\frac{\pi\eta_D^2}{\omega_1}\bigg|+O(\eta^3).
\end{equation}
Eq.(\ref{action2}) can be compared with the exact result for the square
(\ref{squareaction}) where with $e_1=-e_2=1,~e_3=0$ we have 
\begin{equation}
\omega_1 = -i\omega_2=
\frac{\sqrt{\pi}\Gamma({\frac{1}{4}})}{4\Gamma(\frac{3}{4})}. 
\end{equation}
Using
\begin{equation}
\eta_D(i) =\frac{\Gamma(\frac{1}{4})}{2\pi^{3/4}}~,~~~~~~~~
\frac{\Gamma({\frac{1}{4}})\Gamma({\frac{3}{4}})}{\pi}=\sqrt{2}
\end{equation}
we obtain
\begin{equation}
\frac{1}{2\pi}S_z =- \eta\log\eta+ 
\eta \big(1-2\log\frac{\Gamma(\frac{3}{4})}{\Gamma(\frac{1}{4})}-
3 \log 2\big)+\eta^2 \log 2 
\end{equation}
which agrees with the expansion to second order of (\ref{squareaction}).
Similarly one compares eq.(\ref{action2}) with the exact result
(\ref{equirhombusaction}) for the equianharmonic case finding agreement.  

Using the exact relation
\begin{equation}
\frac{1}{2}\partial^2_z \phi -\frac{1}{4}(\partial_z \phi)^2=\epsilon
(\beta+\wp(z)) 
\end{equation}
and the expression of $\phi_1$ (\ref{phi1}) one can also retrieve the 
accessory parameter
$\beta$ to first order i.e. eq.(\ref{betafirst}) using 
\begin{equation}
\zeta(z)=\frac{\zeta(\omega_1)z}{\omega_1}+\frac{\pi}{2\omega_1}
\frac{\theta_1'(\frac{\pi z}{2\omega_1}|\tau)}{\theta_1(\frac{\pi
z}{2\omega_1}|\tau)}
\end{equation}
and the Legendre relation \cite{batemanII}. 

It is easy in this framework to obtain $\phi_2(0)$ i.e. $X$ to second order
which integrated provides the action to third order in $\eta$ for any $\tau$.
This would be very tedious to obtain in the approach described in section
\ref{secondperturbation} which however provides the value of the accessory
parameter to second order.

We must add to eq.(\ref{X01}) $\phi_2(0)$ given by
\begin{equation}
\phi_2(0)=-\frac{16\pi^2\eta^2}{\pi^2\omega_1^2}\int G(z')G(-z') dz\wedge
d\bar z\frac{i}{2}=
-\frac{4\eta^2\tau_2^2}{\pi^2}{\sum_{mn}}'\frac{1}
{(m-\tau n)^2(m-\bar\tau n)^2}
\end{equation}
where the prime means $m=n=0$ excluded. Such a sum, using standard resummation
formulas \cite{weil}, can be rewritten in terms of two simple 
sums which converge
rapidly due to 
the presence of the imaginary part of $\tau$
\begin{eqnarray}
s={\sum_{mn}}'\frac{1}
{(m-\tau n)^2(m-\bar\tau n)^2} =\frac{\pi^4}{45}+
2\pi{\sum}_n'\frac{1}{n^3(\tau-\bar\tau)^3}(\cot \pi\tau n-\cot \pi\bar\tau
n)\nonumber\\
+\pi^2{\sum_n}'\frac{1}{n^2(\tau-\bar\tau)^2}(\frac{1}{\sin^2\pi\tau n}+
\frac{1}{\sin^2\pi\bar\tau n}).
\end{eqnarray}

If we integrate in $\eta$ according to eq.(\ref{Xrelation}) we obtain the 
third order
contribution to the one-point function
\begin{eqnarray}
\frac{1}{2\pi}S^{(3)}_z = \frac{4\eta^3\tau_2^2}{3\pi^2}\eta^3 s~.
\end{eqnarray}

\bigskip

\section{General analytic properties}\label{analyticproperties}

In this section we shall examine the general analytic properties of the
accessory parameter as a function of the coupling and of the modulus 
at the non perturbative level. 

We start with the remark that the uniqueness of Picard solution implies also
the uniqueness of the accessory parameter as
\begin{equation}
e^{\varphi/2}\partial^2_ue^{-\varphi/2} = -Q(u)
\end{equation}
which identifies uniquely $\beta$. Actually $\beta$ can be obtained from 
\begin{equation}\label{contourbeta}
\frac{1}{2\pi i}\oint_{e_1} e^{\varphi/2}\partial^2_ue^{-\varphi/2}du=
\frac{\epsilon}{4}\frac{e_1+\beta}{(e_1-e_2)(e_3-e_1)}+
\frac{3e_1}{8(e_1-e_2)(e_1-e_3)}~.
\end{equation}
In \cite{CMS2} it was proven using Green function technique
that when $\epsilon$ varies in the physical interval 
$0<\epsilon\leq 1/4$, the functions
$\varphi$, $\partial_u\varphi$, $\partial^2_u\varphi$ are 
uniformly
bounded functions of $u$  in any region of the $u$ plane, 
obtained by excluding finite disks around the singularities, with bounds which
depends continuously on $\epsilon$.

Thus taking
the contour of the integral (\ref{contourbeta}) at a finite distance 
from $e_1$ we have that
$\beta$ is a bounded function of $\epsilon$ when it varies in the physical
region. Such a result combined with the uniqueness of the solution implies
that $\beta$ is a continuous function of $\epsilon$. In fact if $\epsilon_n$ is
a sequence of values converging to $\epsilon_0$, due to the boundedness the
corresponding sequence $\beta_n$ must have at least one limit
point. However a limit point due to the continuity of the basic relations 
(\ref{mastersphere},\ref{mastercycle}) is a 
solution of the monodromy problem and being such solution unique there must be
only one limit point. Continuity plays an important role in the following as
in most of the procedures related to the zeros of analytic functions
\cite{whitney}.
 
Starting from the relation (\ref{mastercycle}) we recall that if at a point
$\epsilon_0$ in the physical range $M_{12}(C_2)\neq 0$ we have also
$M_{21}(C_2)\neq 0$ as explained in section \ref{differentialequation}. 
On the other hand if
$M_{12}(C_2)=0$ we have also $M_{21}(C_2)=0$. We cannot have at the same time
$M_{12}(C_1)=0$ and $M_{12}(C_2)=0$ otherwise the parameter $\kappa$ would be
left undetermined against Picard's uniqueness theorem. Thus given any value
$\epsilon_0$ in the physical region dividing either by 
$M_{12}(C_2)\overline{M_{21}(C_2)}$ or by $M_{12}(C_1)\overline{M_{21}(C_1)}$ 
we reach in an open interval around $\epsilon_0$ the structure
\begin{equation}
A(\beta,\epsilon)=\bar B(\bar\beta,\epsilon)
\end{equation} 
with $A$ analytic function of  $\beta$ and $\epsilon$ and $\bar B$ analytic 
function of $\bar\beta$ and $\epsilon$.

As done in section \ref{secondperturbation} it will be useful to employ 
the polarization technique \cite{dangelo} introducing in
addition to $\beta$ an other independent complex variable  $\beta_c$.

We consider now the system 
\begin{eqnarray}\label{Esystem}
E_1 = A(\beta,\epsilon)-\bar B(\beta_c,\epsilon)=0 \nonumber\\
E_2 =B(\beta,\epsilon)-\bar A(\beta_c,\epsilon)=0.
\end{eqnarray}
We look for solutions of the above system for $0<\epsilon\leq 1/4$. Obviously
if $(\beta, \beta_c)$ is a solution also $(\bar\beta_c,\bar\beta)$ 
is a solutions
but we shall be particularly interested in self-conjugate solutions i.e. those
for which $\beta_c=\bar\beta$ insofar they are the solution of the monodromy
problem. Actually from the existence and uniqueness result of the
monodromic solution we know that for $0<\epsilon\leq 1/4$ there is always one
and only one self-conjugate solution, in addition, possibly, to other non
self-conjugate solutions. In the following we shall denote such unique
self-conjugate solution as $\beta(\epsilon)$.

For $\epsilon=\epsilon_0$ we have 
\begin{equation}
A(\beta(\epsilon_0),\epsilon_0)=\bar B(\overline{\beta(\epsilon_0)},
\epsilon_0)~.
\end{equation}
The Weierstrass preparation theorem \cite{gunningrossi,whitney} can be applied
to $A(\beta,\epsilon)-A(\beta(\epsilon_0),
\epsilon_0)$ if 
\begin{equation} 
A(\beta,\epsilon_0)-A(\beta(\epsilon_0),\epsilon_0)
\end{equation} 
is not identically zero in $\beta$. This can happen only at a finite number of
points in the open interval otherwise
$\frac{\partial A}{\partial \beta}\equiv 0$ 
i.e. $A$ would be a function only of $\epsilon$ which from the structure of
the $M_{jk}(C_l)$ of section 3 is not true.
We exclude such a finite number of points.

Thus except at most a finite number of points we can apply 
Weierstrass preparation
theorem \cite{gunningrossi,whitney} 
\begin{equation}\label{weierA} 
A(\beta,\epsilon)-A(\beta(\epsilon_0),\epsilon_0)
=u(\beta,\epsilon)((\beta-\beta(\epsilon_0))^m+
c_{m-1}(\epsilon) (\beta-\beta(\epsilon_0))^{m-1}+ ... 
+c_0(\epsilon))
\end{equation}
\begin{equation}\label{weierbB}
\bar B(\beta_c,\epsilon)-\bar B(\overline{\beta(\epsilon_0)},\epsilon_0)
=v(\beta_c,\epsilon)((\beta_c-\overline{\beta(\epsilon_0)})^n+
g_{n-1}(\epsilon) (\beta_c-\overline{\beta(\epsilon_0)})^{n-1}+ ... 
+g_0(\epsilon))
\end{equation} 
with $u(\beta,\epsilon), v(\beta_c,\epsilon)$ units and  $c_k,g_k$ analytic
functions of $\epsilon$, vanishing at $\epsilon_0$. 

We consider first the case: $m=n=1$.

At $\epsilon_0,\beta(\epsilon_0),\overline{\beta(\epsilon_0)}$ we have for the
system (\ref{Esystem}) the Jacobian 
\begin{equation}
J=\frac{\partial(E_1,E_2)}{\partial(\beta,\beta_c)}=
-|u(\beta(\epsilon_0),\epsilon_0)|^2+ |v(\overline{\beta(\epsilon_0)},
\epsilon_0)|^2~.
\end{equation}
If $J\neq 0$ we can apply the  implicit
function theorem according to which the solution $\beta$ and $\beta_c$ is
unique (and thus self-conjugate for real $\epsilon$) and $\beta$ is an analytic
function in an open interval around $\epsilon_0$ and thus we have local
analyticity.

If $J=0$ then we look at the equation
\begin{equation}\label{Jzero}
u(re^{i\alpha}+\beta(\epsilon_0),\epsilon_0)~ r e^{i\alpha}-
v(re^{-i\alpha}+\overline{\beta(\epsilon_0)},\epsilon_0)~ r e^{-i\alpha}=0.
\end{equation}
For $r\neq 0$  divide (\ref{Jzero}) by
$r$ and call it $F(r,\alpha)$. 
\begin{equation}
F(r,\alpha)=u(re^{i\alpha}+\beta(\epsilon_0),\epsilon_0)e^{i\alpha}
-v(re^{-i\alpha}+\overline{\beta(\epsilon_0)},\epsilon_0)e^{-i\alpha}=0~.
\end{equation}
Consider a solution $\alpha_0$ of $F(0,\alpha)=0$
\begin{equation}
u(\beta(\epsilon_0),\epsilon_0) e^{i\alpha_0}= 
v(\overline{\beta(\epsilon_0)},\epsilon_0) e^{-i\alpha_0}
\end{equation} 
which is soluble because $J=0$.
Then in the product of the open intervals
$\alpha_0-\delta<\alpha<\alpha_0+\delta$, $-\delta_r< r<\delta_r$
we have that $F(r,\alpha)$ is a $C^1$ function of $\alpha$ and $r$,  
with $F(0,\alpha_0)=0$ and 
$F_\alpha(0,\alpha_0)\neq 0$. Then for small $r$ 
we have one 
solution $\alpha(r)$ for $\alpha$ \cite{rudin}, thus a self-conjugate 
solution with $r\neq
0$ (any $r$ in the above interval) in addition to the
$\beta=\beta(\epsilon_0),\beta_c=\overline{\beta(\epsilon_0))}$. This however
violates Picard's uniqueness result. 
The conclusion is that either $J\neq 0$ or the Weierstrass polynomials 
(\ref{weierA},\ref{weierbB}) cannot be both first order.

In the same way one excludes Weierstrass polynomials (\ref{weierA}) 
and (\ref{weierbB}) with the
same order, $m=n >1$ and $|u(\beta(\epsilon_0),\epsilon_0)|
=|v(\overline{\beta(\epsilon_0)},\epsilon_0)|$.

\bigskip

We go back now to the system (\ref{Esystem})

Given $\epsilon_0$ we have  
\begin{eqnarray}
\bar B(\overline{\beta(\epsilon_0)},\epsilon_0)
-A(\beta(\epsilon_0),\epsilon_0)=0
\nonumber\\
\bar A(\overline{\beta(\epsilon_0)},\epsilon_0)-
B(\beta(\epsilon_0),\epsilon_0)=0
\end{eqnarray}
and in a neighborhood $\Delta_0$ of $\epsilon_0$, $\Delta_\beta$ of 
$\beta(\epsilon_0)$, 
$\Delta_{\beta c}$ of $\overline{\beta(\epsilon_0)}$ using 
Weierstrass preparation theorem
we can write system (\ref{Esystem}) as
\begin{eqnarray}
U(\beta,\beta_c,\epsilon) P_1(\beta_c-\overline{\beta(\epsilon_0)}
;\beta,\epsilon)=0 \nonumber\\
V(\beta,\beta_c,\epsilon) P_2(\beta_c-\overline{\beta(\epsilon_0)}
;\beta,\epsilon)=0
\end{eqnarray}
which, as $U$ and $V$ are units, is equivalent to
\begin{eqnarray}\label{master3}
P_1(\beta_c-\overline{\beta(\epsilon_0)}
;\beta,\epsilon)=0 \nonumber\\
P_2(\beta_c-\overline{\beta(\epsilon_0)}
;\beta,\epsilon)=0~.
\end{eqnarray}

Necessary and sufficient condition for the two polynomials in (\ref{master3}) 
to have a common solution in $\beta_c$ is that the resultant 
\cite{vdW,whitney,lang} 
of the two polynomials $P_1$ and $P_2$ is zero
\begin{equation}\label{fequation}
R(P_1,P_2)\equiv f(\beta,\epsilon)=0~.
\end{equation} 
In particular we know from the existence result that
\begin{equation}
f(\beta(\epsilon_0),\epsilon_0)=0~.
\end{equation} 
Exploiting again Weierstrass preparation theorem eq.(\ref{fequation}) 
can be written for 
$\epsilon\in\Delta_1\subset\Delta_0$, $\beta\in \Delta_{\beta 1}
\subset \Delta_\beta
$ as
\begin{equation}
u(\beta,\epsilon)~P(\beta-\beta(\epsilon_0);\epsilon)=0
\end{equation} 
with
\begin{equation}\label{betaepsWpoly}
P(\beta-\beta(\epsilon_0);\epsilon)= (\beta-\beta(\epsilon_0))^m+
(\beta-\beta(\epsilon_0))^{m-1} a_{m-1}(\epsilon)+\cdots+a_0(\epsilon).
\end{equation}
In order to apply Weierstrass preparation theorem to $f(\beta,\epsilon)$
we need that $f(\beta,\epsilon_0)$ does not vanish identically in $\beta$.
The vanishing of $f(\beta,\epsilon_0)$ would mean that the system 
(\ref{Esystem}) at $\epsilon_0$ has solution for all $\beta$ near
$\beta(\epsilon_0)$. 
This means, using the Weierstrass-polynomial expression for 
$A$ and $\bar B$, that $m=n$ and $|u(\beta(\epsilon_0),\epsilon_0)|=
|v(\overline{\beta(\epsilon_0)},\epsilon_0)|$ and this implies the existence 
of infinite self-conjugate solutions with $\beta\neq \beta(\epsilon_0)$ 
at $\epsilon_0$ and this goes against the uniqueness theorem.
				       
We start now by computing the resultant $R(P,P')$ i.e. the discriminant of
$P$. If it is
not identically zero it can vanish in the interval around $\epsilon_0$
included in the Weierstrass set at most at a finite number of points,
otherwise it would be identically zero.  Thus except at those finite number of
points we can apply the analytic implicit function theorem \cite{gunningrossi}
to have analyticity of $\beta(\epsilon)$ in a open interval around
$\epsilon_0$.

The general case can be treated by computing the reduced Gram determinants
$D_n$ of the power-vectors of the roots \cite{whitney}
\begin{equation}
D_n=
\begin{vmatrix}
s_0&s_1&\cdots&s_{n-1}\\
s_1&s_2&\cdots&s_n\\
\cdots&\cdots&\cdots&\cdots\\
s_{n-1}&s_n&\cdots&s_{2n-2}
\end{vmatrix}
\end{equation} 
where
\begin{equation}
s_i = \xi_1^i+\xi_2^i+\cdots+\xi_m^i
\end{equation}
being $\xi_k$ the $m$ roots of $P$. Being $D_n$ a symmetric polynomial of the
roots it is a polynomial in the coefficients $a_k(\epsilon)$ and as such an
analytic function of $\epsilon$.
If $R(P,P')\equiv D_m$ vanishes identically it means that we have at each
$\epsilon$ a double or higher order root.
Then compute $D_{m-1}$. If it is not identically zero it means that the
maximum number of distinct roots is $m-1$ and the set where they are $m-1$
is open and given by subtracting from the initial open set the zeros of
$D_{m-1}$. These  are isolated points \cite{whitney} and thus 
finite in number. 
Moreover in the region where the maximum number
of distinct roots is reached all the solutions of
$P(\beta-\beta(\epsilon_0);\epsilon)=0$ (the so called local sheets) 
are analytic \cite{whitney}, and in particular Picard solution is analytic.

Suppose now that $D_m=D_{m-1}\equiv 0$. Then compute $D_{m-2}$ and proceed 
as above. If $D_{m-2}$ is not identically zero it means that the
maximum number of distinct roots is $m-2$; it can vanish only at a finite
number of point and except at those points all solutions of (\ref{fequation}) 
are analytic.

The procedure ends due to the fact that $D_1=s_0\equiv m$. The vanishing of all
$D_n$ , $n>1$  corresponds
to the situation where we have only one $m$-time degenerate solution
i.e.
\begin{equation}
P(\beta-\beta_0;\epsilon) = (\beta-\beta(\epsilon))^m
\end{equation}
from which $\beta(\epsilon)-\beta(\epsilon_0)
=-\frac{1}{m}a_{m-1}(\epsilon) $ which is analytic.

Removing the described 
finite number of points we have that given any $\epsilon_0$ there is an open
disk around $\epsilon_0$ where all the solutions of (\ref{fequation}) and in
particular the unique self-conjugate Picard solution, are analytic except for
at most a finite number of points .

We saw in section \ref{secondperturbation}
that a finite interval around the origin $0<\epsilon\leq \delta$   
is covered by the convergent perturbation theory treatment. 
For the remainder $\delta\leq\epsilon\leq 1/4$ we can associate to each
$\epsilon$ an open set with the above properties and then as
$\delta\leq\epsilon\leq 1/4$ is compact we can extract a finite covering.
We conclude that the unique self-conjugate Picard solution is analytic on the
whole physical region except at most at a finite number of points.

\bigskip

Similarly one treats the dependence of $\beta$ on the modulus $\tau$.

Choose any $\tau_0$ belonging to the fundamental region and $\epsilon_0$ with
$0<\epsilon_0\leq 1/4$. 

From now on we shall neglect in the notation $\epsilon_0$ i.e. we
shall work at fixed $\epsilon$.

We start again from the equation 
\begin{equation}
A(\beta,\tau)-\bar B(\bar\beta,\bar\tau)=0. 
\end{equation}

\bigskip

As done above it is useful to apply the polarization technique to $\beta$ by
introducing an
other independent complex variable $\beta_c$, but
this time we apply the polarization technique also to the variable $\tau$, the
modulus, by introducing in addition to $\tau$ an independent complex variable
$\tau_c$. We remark that in the previous treatment of the dependence of
$\beta$ on $\epsilon$ we could have applied the polarization technique also to
the variable $\epsilon$ but being the physical values of
$\epsilon$ real we would have reached the same results. Here instead the
physical values of $\tau$ are in the complex.

We consider the system 
\begin{eqnarray}\label{tausystem}
A(\beta,\tau)-\bar B(\beta_c,\tau_c)&=&0 \nonumber\\
B(\beta,\tau)-\bar A(\beta_c,\tau_c)&=&0~.
\end{eqnarray}
We look for solutions of the above system for $\tau$ in the fundamental region
and $\tau_c=\bar\tau$. Obviously
if $\beta, \beta_c$ is a solution also $\bar\beta_c,\bar\beta$ is a solutions
but we shall be interested is self-conjugate solutions i.e. those
for which $\beta_c=\bar\beta$ insofar they are the solution of the monodromy
problem. Actually from the existence and uniqueness result of the
monodromic solution we know that for $0<\epsilon_0\leq 1/4$ and
$\tau_c=\bar\tau$ there is always one
and only one self-conjugate solution, in addition, possibly, to other non
self-conjugate solutions. In the following we shall denote the unique
self-conjugate solution as $\beta(\tau)$.

Chosen $\tau_0$ in the fundamental region we have 

\begin{equation}
A(\beta(\tau_0),\tau_0)=\bar B(\overline{\beta(\tau_0)},\bar\tau_0)~.
\end{equation}
Applying the Weierstrass preparation theorem to
$A(\beta,\tau)$ and $\bar B(\beta_c,\tau_c)$
we have 
\begin{equation}\label{weiertauA}
A(\beta,\tau)-A(\beta(\tau_0),\tau_0)
=u(\beta,\tau)((\beta-\beta(\tau_0))^m+
c_{m-1}(\tau) (\beta-\beta(\tau_0))^{m-1}+ ... 
+c_0(\tau))
\end{equation}
\begin{equation}\label{weiertaubB}
\bar B(\beta_c,\tau_c)-\bar B(\overline{\beta(\tau_0)},\bar\tau_0)
=v(\beta_c,\tau_c)((\beta_c-\overline{\beta(\tau_0)})^n+
g_{n-1}(\tau_c) (\beta_c-\overline{\beta(\tau_0)})^{n-1}+ ... 
+g_0(\tau_c))
\end{equation}
with $c_k$ analytic functions of $\tau$ vanishing at $\tau_0$ and $g_k$
analytic functions of  $\tau_c$ vanishing at $\bar\tau_0$ and $u$ and $v$
units.

As done for the dependence on $\epsilon$, for $m=n=1$
if $J(\tau_0,\bar\tau_0)\neq 0$ we are in the analytic situation while 
$J(\tau_0,\bar\tau_0)=0$ is excluded by the uniqueness result. 
In the same way one excludes Weierstrass polynomials (\ref{weiertauA}) 
and (\ref{weiertaubB}) with the
same order $m=n> 1$ and $|u(\beta(\tau_0),\tau_0)|
=|v(\overline{\beta(\tau_0)},\bar\tau_0)|$.

Given $\tau_0$ as 
\begin{eqnarray}
\bar B(\overline{\beta(\tau_0)},\bar\tau_0)-A(\beta(\tau_0),\tau_0)=0
\nonumber\\
\bar A(\overline{\beta(\tau_0)},\bar\tau_0)-
B(\beta(\tau_0),\tau_0)=0
\end{eqnarray}
in a neighborhood $\Delta$ of $\tau_0$, $\Delta_{c}$ of $\bar\tau_0$, 
$\Delta_\beta$ of $\beta(\tau_0)$, 
$\Delta_{\beta c}$ of $\overline{\beta(\tau_0)}$ we can write system 
(\ref{tausystem}) as
\begin{eqnarray}
U(\beta,\beta_c,\tau,\tau_c) P_1(\beta_c-\overline{\beta(\tau_0)}
;\beta,\tau,\tau_c)=0 \nonumber\\
V(\beta,\beta_c,\tau,\tau_c) P_2(\beta_c-\overline{\beta(\tau_0)}
;\beta,\tau,\tau_c)=0
\end{eqnarray}
which as $U$ and $V$ are units is equivalent to
\begin{eqnarray}\label{Psystem}
P_1(\beta_c-\overline{\beta(\tau_0)}
;\beta,\tau,\tau_c)&=&0 \nonumber\\
P_2(\beta_c-\overline{\beta(\tau_0)}
;\beta,\tau,\tau_c)&=&0~.
\end{eqnarray}

A common solution of (\ref{Psystem}) in $\beta_c$ implies (necessary and
sufficient condition)  
the resultant of $P_1,P_2$ to be zero
\begin{equation}\label{fbetatautauc}
R(P_1,P_2)\equiv f(\beta,\tau,\tau_c)=0~.
\end{equation}
In particular we know from the existence result that
\begin{equation}
f(\beta(\tau_0),\tau_0,\bar\tau_0)=0.
\end{equation}
$f(\beta,\tau_0,\bar\tau_0)$ cannot be identically zero in $\beta$ for the
same reasoning as the one performed after eq.(\ref{betaepsWpoly}); 
thus we can apply
Weierstrass preparation theorem and write
$f(\beta,\tau,\tau_c)$ for 
$\tau\in\Delta_1\subset \Delta$, 
$\tau_c\in\Delta_{1c}\subset \Delta_c$, 
$\beta\in \Delta_{\beta 1}\subset \Delta_\beta$ as
\begin{equation}
f(\beta,\tau,\tau_c)= u(\beta,\tau,\tau_c)P(\beta-\beta(\tau_0);
\tau,\tau_c)~~~~
\end{equation}
with
\begin{equation}\label{betaepsW}
P(\beta-\beta(\tau_0);\tau,\tau_c)= (\beta-\beta(\tau_0))^m+
a_{m-1}(\tau,\tau_c)(\beta-\beta(\tau_0))^{m-1}+
\dots+a_0(\tau,\tau_c)~~~~
\end{equation}
and the coefficients $a_n(\tau,\tau_c)$ analytic in $\tau,\tau_c$ and
vanishing at $\tau_0,\bar\tau_0$.

Thus the equation has become
\begin{equation}
P(\beta-\beta(\tau_0);\tau,\tau_c)=0.
\end{equation}
$\beta$ is analytic in $\tau,\tau_c$ at all points $\tau,\bar\tau$ except
those at which 
$P'(\beta(\tau);\tau,\bar\tau)=0$. These $\tau$ satisfy  
the discriminant equation
\begin{equation}
R(P,P')\equiv D_m(\tau,\bar\tau)=0
\end{equation}
with $R(P,P')$ analytic in $\tau,\tau_c$ being a polynomial in the
$a_n(\tau,\tau_c)$.  

We distinguish two cases

1. $D_m(\tau,\bar\tau)$ is identically zero. Then due to a theorem on
polarization \cite{dangelo} $D_m(\tau,\tau_c)$ is identically zero.
 
2. Otherwise $D_m(\tau,\tau_c)$ can vanish only on a {\it thin} set
\cite{gunningrossi,whitney}, of which the points $\tau$ such that
$D_m(\tau,\bar\tau)=0$ are a subset. Thin set have zero measure
\cite{gunningrossi}. Outside such thin
set the equation is invertible and thus $\beta$ analytic function of
$\tau,\tau_c$, i.e. $\beta(\tau,\bar\tau)$ a real-analytic function of $\tau$. 

In case 1 i.e. 
\begin{equation}
D_m(\tau,\tau_c)\equiv 0
\end{equation}
we compute $D_{m-1}$. If it is not identically zero it means that the 
maximum number of distinct roots is $m-1$ and the set where they are $m-1$ 
is open and given by subtracting from the initial open set the zeros of 
$D_{m-1}$ which is a thin set and as such of zero measure. 
In the region where the maximum number
of distinct roots is reached all the solutions of (\ref{fbetatautauc}) 
(local sheet) are analytic \cite{whitney}, and in particular Picard solution is
analytic.  

Suppose now that 
\begin{equation}
D_m=D_{m-1}\equiv 0~.
\end{equation}
Then we compute $D_{m-2}$ an proceed as above. 

The procedure ends due to the fact that $D_1\equiv m$. It corresponds
to the situation where we have only one $m$-times degenerate solution
i.e.
\begin{equation}
P(\beta-\beta(\tau_0);\tau,\tau_c) = (\beta-\beta(\tau,\tau_c))^m =0
\end{equation}
from which we have
$\beta(\tau,\tau_c)-\beta(\tau_0)=-
\frac{1}{m}a_{m-1}(\tau,\tau_c) $ which is analytic in $\tau,\tau_c$ and thus
$\beta(\tau,\bar\tau)$ real-analytic in $\tau$.

We can divide the fundamental region of $\tau$ in a denumerable set of
horizontal strips which are compact. We have a zero-measure set of possible
non real-analyticity points in each strip and the union of such infinite
zero measure set has zero measure.

We conclude that for each $\epsilon$ in the physical region the accessory
parameter $\beta$ is a real-analytic function of $\tau$ in the whole
fundamental region except at most for a zero measure set.

\section{Conclusions}\label{conclusions}

We have considered the problem of accessory parameters on the torus. 
The specific case we dealt with is that of a
single source which corresponds to a special cases of the Heun equation. We
proved that necessary and sufficient condition to obtain monodromy at all
singularities is the fulfillment of a single implicit equation. Several
features of the accessory parameter can be extracted from such an equation.
A perturbative series was developed and rigorous
lower bound on the radius of convergence of the perturbative series has been
given. The second order result for the accessory parameter and 
third order result for the one point function was explicitly computed.  

Modular invariance
is useful to find the value of the accessory parameter in some special cases
and it is satisfied by the perturbative solution. 
General analytic properties of
the dependence of the accessory parameter on the source strength and on the
modulus have been proved. The real-analyticity of the dependence of
the accessory parameters on the moduli is an essential step in proving
Polyakov relation on the sphere which has the meaning of determining the
response of the on-shell action on the position of the singularities.
We shall devote a separate paper to the structure and meaning of the Polyakov
relation on the torus.

The described technique can be extended to treat the four-point case
or higher number of points on the sphere, higher point function on the
torus or higher genus surfaces.

In \cite{ferraripiatek} an integral expression has been given for the accessory
parameter for the four point function on the sphere.  Such a procedure should
be extensible to the one point function on the torus. However
comparison 
of that result e.g. with the second and third order result of sections
\ref{secondperturbation}, \ref{phiperturbation} will not
be immediate as our result is an expansion in the source strength while the
results of \cite{ferraripiatek} are nearer to an expansion in the position of
the singularities which in the case of the torus correspond to the
value of the modulus.

On the other hand in \cite{torusI,torusII} analytical technique for dealing
with the deformation of the torus have been developed. These could allow the
comparison of the result obtained along the lines of \cite{ferraripiatek}.




\begin{thebibliography}{99}

\bibitem{curtrightthorn} T.L. Curtright, C.B. Thorn, Phys. Rev. Lett. 48 (1982)
1309; Erratum-ibid. 48 (1982) 1768

\bibitem{dornotto} H. Dorn, H.J. Otto, Nucl. Phys. B429 (1994) 375, 
e-Print arXiv:hep-th/9403141

\bibitem{teschner} J. Teschner, Class. Quant. Grav.18 (2001) R153, 
e-Print arXiv:hep-th/0104158; Int. J. Mod. Phys. A19S2 (2004) 436, 
e-Print arXiv:hep-th/0303150; 
Phys. Lett. B363 (1995) 65, e-Print arXiv:hep-th/9507109
 
\bibitem{ZZ} A.B. Zamolodchikov, Al.B. Zamolodchikov, 
Nucl. Phys. B477 (1996) 577, e-Print arXiv:hep-th/9506136; e-Print
arXiv:hep-th/0101152 

\bibitem{olesen} P. Olesen, Phys. Lett. B265 (1991) 361;
Phys. Lett. B268 (1991) 389

\bibitem{jackiwpi} R. Jackiw, S.Y. Pi, Phys. Rev. Lett. 64 (1990) 2969

\bibitem{akerblom} N. Akerblom, G. Cornelissen, G. Stavenga, J.-W. van Holten,
J.Math.Phys. 52 (2011) 072901, e-Print arXiv:0912.0718 [hep-th]

\bibitem{gubser} S.S. Gubser, I.R. Klebanov, A.M. Polyakov, Phys. Lett. B428
(1998) 105, e-Print arXiv:hep-th/9802109

\bibitem{nakayama} Y. Nakayama, Int. J. Mod. Phys. A19 (2004) 2771, e-Print
arXiv:hep-th/0402009

\bibitem{AGT} L.F. Alday, D. Gaiotto, Y. Tachikawa,
Lett. Math. Phys. 91 (2010) 167, e-Print arXiv:0906.3219 [hep-th]

\bibitem{gaiotto}D. Gaiotto, e-Print arXiv:0904.2715 [hep-th]

\bibitem{drukker} N. Drukker, J. Gomis, T. Okuda, J. Teschner, JHEP
1002:057,2010, e-Print arXiv:0909.1105 [hep-th]

\bibitem{alba}V. Alba, A. Morozov, JETP Lett. 90 (2009) 708-712,
e-Print arXiv:0911.0363 [hep-th]

\bibitem{hadasz1} L. Hadasz, Z. Jaskolski, P. Suchanek, JHEP 1006:046
(2010), e-Print arXiv:1004.1841 [hep-th]

\bibitem{hadasz2} L. Hadasz, Z. Jaskolski, P. Suchanek, JHEP
1001:063,2010, e-Print arXiv:0911.2353 [hep-th]; Phys. Lett. B685 (2010) 79,
e-Print arXiv:0911.4296 [hep-th]

\bibitem{MV} P. Menotti, G. Vajente, Nucl. Phys. B709 (2005) 465,
e-Print arXiv:hep-th/0411003

\bibitem{FLNO} V.A. Fateev, A.V. Litvinov, A. Neveu, E. Onofri,
J.Phys. A42:304011 (2009), e-Print arXiv:0902.1331 [hep-th]

\bibitem{ferraripiatek} F. Ferrari, M. Piatek, JHEP 1205 (2012) 025,
e-Print arXiv:1202.2149 [hep-th]

\bibitem{KRV} L. Keen, H.E.Rauch,  A.T. Vasquez, Trans.Am. Math.Soc.255 (1979)
201

\bibitem{torusI} P. Menotti, J.Phys. A44:115403 (2011), e-Print
arXiv:1010.4946 [hep-th]

\bibitem{torusII} P. Menotti, J.Phys. A44:335401 (2011),
e-Print arXiv:1104.3210 [hep-th]

\bibitem{maier} R.S. Maier, J.Differential Equations, 213 (2005) 171

\bibitem{picard} E. Picard, Compt.Rend. 116 (1893) 1015; J.Math.Pures
Appl. 4 (1893) 273 and (1898) 313; Bull.Sci.Math. XXIV 1 (1900) 196

\bibitem{poincare} H. Poincar\'e,  J. Math. Pures Appl. (5) 4 (1898) 137

\bibitem{lichtenstein} L. Lichtenstein, Acta Mathematica 40 (1915) 1

\bibitem{troyanov}  M. Troyanov, Trans. Am.Math.Soc. 324 (1991) 793

\bibitem{ZT} P. G. Zograf, L. A. Takhtajan, Math. USSR Sbornik 60 (1988)
143 

\bibitem{kra} I. Kra, Trans. Am.Math.Soc. 313 (1989) 589

\bibitem{CMS1} L. Cantini, P. Menotti, D. Seminara, Phys.Lett. B517 (2001)
203, e-Print hep-th/0105081

\bibitem{CMS2} L. Cantini, P. Menotti, D. Seminara, Nucl.Phys. B638 (2002)
351, e-Print hep-th/0203103 

\bibitem{whitney} H. Whitney ``Complex analytic varieties'' Addison-Wesley,
Reading Mass. (1972)

\bibitem{TZ} L. A. Takhtajan, P. G. Zograf Takhtajan, Trans. Am.
Math. Soc. 355 (2003) 1857   

\bibitem{batemanII} A. Erdelyi (Ed.) ``Higher Transcendental Functions'', 
vol.II McGraw-Hill, New York (1953)

\bibitem{DLMF} NIST Digital Library of Mathematical Functions:
http://dlmf.nist.gov/ 

\bibitem{gunningrossi} R. C. Gunning, H. Rossi ``Analytic functions of several
complex variables'', Prentice-Hall Inc.Englewood Cliffs (1965)

\bibitem{dangelo} J. P. D'Angelo ``Several complex variables and the geometry
of real hypersurfaces'' CRC Press, Ann Arbor, London, Tokio (1993) 

\bibitem{weil} A. Weil ``Elliptic Functions According to
Eisenstein and Kronecker'' Springer, Berlin (1976)

\bibitem{rudin} W. Rudin ``Principles of mathematical analysis'' McGraw-Hill,
New York (1976)

\bibitem{vdW} B.L. van der Waerden ``Algebra'' Springer-Verlag, New
York, Heidelberg, Berlin (1967) 

\bibitem{lang}  S. Lang ``Algebra'' Addison-Wesley,
Reading Mass. (1993)

\bibitem{takhtajan} L.A. Takhtajan, Mod.Phys.Lett. A11 (1996) 93,
e-Print hep-th/9509026 [hep-th] 

\bibitem{serre} J-P Serre ``A course in arithmetic'' Springer-Verlag, New
York, Heidelberg, Berlin (1996) {serre}

\bibitem{titchmarsh} E.C. Titchmarsh ``The theory of functions'' Oxford
University Press, London (1964)

\end{thebibliography}
\end{document}